\documentclass[twocolumn,showpacs,preprintnumbers,amsmath,amssymb,prb,superscriptaddress,longbibliography]{revtex4-2}
\usepackage[dvipdfmx]{graphicx}
\usepackage{dcolumn}
\usepackage{bm}
\usepackage{color} 
\usepackage{hyperref}
\hypersetup{
  colorlinks   = true,
  urlcolor     = blue, 
  linkcolor    = black, 
  citecolor   = blue
}
\usepackage[all]{hypcap}
\usepackage{xspace}

\newcommand{\TN}{$T_{\rm N}$\xspace}

\newcommand{\EF}{$E_{\rm F}$\xspace}

\newcommand{\MnTe}{MnTe\xspace}

\newcommand{\OC}{$\sigma_1(\omega)$\xspace}
\newcommand{\R}{$R(\omega)$\xspace}

\newcommand{\hw}{$\hbar\omega$\xspace}

\newcommand{\Neff}{$N^{*}$\xspace}

\begin{document}   

\title{
Band structure control in the altermagnetic candidate MnTe \\by temperature and strain
}

\author{Shin-ichi~Kimura}
\email{sk@kimura-lab.com}
\affiliation{Graduate School of Frontier Biosciences, The University of Osaka, Suita, Osaka 565-0871, Japan}
\affiliation{Department of Physics, Graduate School of Science, The University of Osaka, Toyonaka, Osaka 560-0043, Japan}
\affiliation{Institute for Molecular Science, Okazaki, Aichi 444-8585, Japan}
\author{Hironao~Suwa}
\affiliation{Department of Physics, Graduate School of Science, The University of Osaka, Toyonaka, Osaka 560-0043, Japan}
\author{Kangle~Yuan}
 \altaffiliation[Present affiliation: ]{The Institute for Solid State Physics, The University of Tokyo, Kashiwa, Chiba 277-8581, Japan}
\affiliation{Department of Physics, Graduate School of Science, The University of Osaka, Toyonaka, Osaka 560-0043, Japan}
\author{Hiroshi~Watanabe}
 \altaffiliation[Present affiliation: ]{Institute for Chemical Research, Kyoto University, Uji, Kyoto 611-0011, Japan}
\affiliation{Graduate School of Frontier Biosciences, The University of Osaka, Suita, Osaka 565-0871, Japan}
\affiliation{Department of Physics, Graduate School of Science, The University of Osaka, Toyonaka, Osaka 560-0043, Japan}
\author{Takuto~Nakamura}
\affiliation{Graduate School of Frontier Biosciences, The University of Osaka, Suita, Osaka 565-0871, Japan}
\affiliation{Department of Physics, Graduate School of Science, The University of Osaka, Toyonaka, Osaka 560-0043, Japan}
\author{Haan Kyul Yun}
\affiliation{Department of Physics, Sogang University, Seoul 04107, Republic of Korea}
\author{Myung-Hwa~Jung}
\affiliation{Department of Physics, Sogang University, Seoul 04107, Republic of Korea}

%

\date{\today}   
\begin{abstract}    
The temperature and strain dependences of the optical conductivity spectrum of hexagonal manganese telluride (MnTe) were measured, revealing absorption in the terahertz (THz) region from spin-split bands to acceptor levels.
The temperature dependence of the THz absorption peak is consistent with that of a ferromagnetic phase transition, even though \MnTe exhibits no net magnetism.
The temperature dependence was attributed to a change in the altermagnetic electronic structure.
Under negative uniaxial strain, the THz peak shifts to the high-energy side, suggesting spin-splitting bands at energies away from \EF, consistent with the theoretical prediction that the spin-splitting angle decreases.
The observed behavior of the THz peak clearly shows that \MnTe has the altermagnetic electronic structure.
Additionally, a Fano-like asymmetric line shape in the optical phonon absorption was observed, possibly originating from interactions with the alternative spin-split bands.
\end{abstract}    



\maketitle


Recently, materials with both antiferromagnetism and time-reversal symmetry broken have attracted attention, namely altermagnet.
The electronic structure with different spin momentum splits below the N\'eel temperature (\TN), consistent with ferromagnetic materials~\cite{Mazin2023-ac}.
The materials have no net magnetism, but owing to the crystal symmetry combined with the antiferromagnetic structure, there are two novel characteristics: a strong time-reversal symmetry-breaking response and the other spin-polarization phenomena typical of ferromagnets~\cite{Smejkal2022-mr}. 
Spin-splitting states in altermagnets are useful for spintronic applications that require fast switching~\cite{Mori2023-wc}.

One candidate for altermagnets is hexagonal manganese telluride (\MnTe) with \TN$\sim 307~{\rm K}$.
The magnetic properties and electronic states of MnTe have been studied for a long time~\cite{Komatsubara1963-nx, Kamat_Dalal1982-wy, Kriegner2017-vv}, and it is known to be an antiferromagnetic semiconductor.
After the material was expected to be an altermagnet, its electronic structure was revealed through angle-resolved photoemission spectroscopy (ARPES)~\cite{Krempasky2024-ew, Lee2024-oe, Osumi2024-uy, Hajlaoui2024-br}, and X-ray magnetic circular dichroism~\cite{Hariki2024-ft}.
ARPES spectra strongly suggest the presence of spin-polarized electronic states characteristic of altermagnets~\cite{Krempasky2024-ew}, and the temperature dependence reveals a peak splitting in the valence band below \TN, which is thought to result from band splitting in the altermagnet~\cite{Lee2024-oe, Osumi2024-uy, Hajlaoui2024-br}.

Optical absorption as well as optical conductivity [$\sigma_1(\omega)$] spectra also have been reported for a long time~\cite{Zanmarchi1967-nm, Allen1977-ae, Angadi1994-nl, Janik1996-dl, Ferrer-Roca2000-fq, Mori2018-lo, Bossini2020-gk}.
Since the primary purposes of the previous works are to investigate the magnitude and the temperature dependence of the energy gap of the semiconducting character.
While ARPES results indicate that the shallowest band approaches the Fermi level (\EF) as temperature decreases, the opposite result is obtained for the absorption edge, which shifts toward higher energies.
This result suggests that not only the occupied spin-split band but also the bottom energy of the unoccupied states change with decreasing temperature; however, this has not been experimentally confirmed, as ARPES can detect the occupied electronic structure below \EF, whereas the unoccupied states cannot be observed. 
\OC spectra contain information about the joint density of states between occupied and unoccupied states. 
From experimental data, the combination of ARPES and \OC spectra provides information on the unoccupied states. 
Additionally, in-gap states with very low density can be observed with \OC spectra.

In addition to temperature-dependent changes in the electronic structure, the band structure of \MnTe has been theoretically predicted to undergo significant changes with strain~\cite{Belashchenko2025-mm}.
By applying these two perturbations of temperature and strain, the altermagnetic band structure is expected to be controlled, thereby enabling band engineering.

In the Letter, we investigate changes in the electronic structure of \MnTe with temperature and strain using \OC spectra over a wide energy range from THz to the vacuum ultraviolet. 
The energy gap, in-gap state, and optical phonon structure are mainly observed and discussed.
Specifically, in the region below 100~meV, optical transitions from the shallowest spin-split band expected in an altermagnet to an acceptor level were observed, exhibiting clear temperature- and strain-dependent behavior.
This result provides evidence that \MnTe is an altermagnet and confirms theoretical predictions.



\begin{figure}
\includegraphics[width=0.4\textwidth]{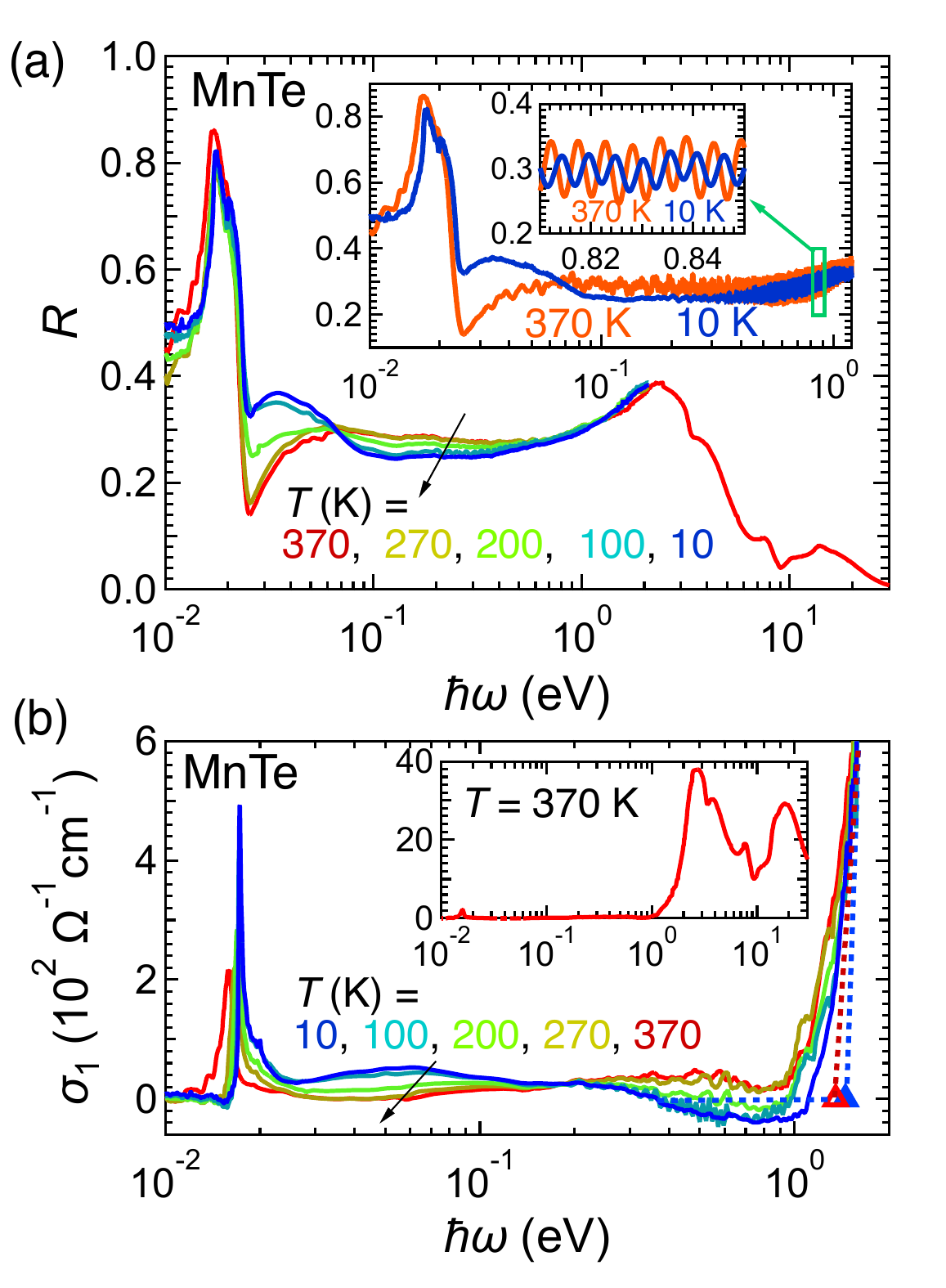}
\caption{
(a) Temperature-dependent reflectivity [$R(\omega)$] spectra of the as-grown $(001)$ surfaces of \MnTe 
in the photon energy \hw range of $0.01-30$~eV after the reduction of the interference shown in the inset.
The interference-reduction method is described in Sec.~S2 in the supplementary material~\cite{SM}.
A peak structure at around \hw $\sim0.02$~eV originates from optical phonons.
(Inset) Raw \R spectra at $T = 10$ and 370~K in the infrared region.
Strong interference is observed in the photon energy above 0.1~eV due to the transparent character of the thin (thickness $\sim0.35$~mm) sample.
(b) Temperature-dependent optical conductivity [$\sigma_1(\omega)$] spectra in the \hw range below 2~eV obtained from the \R spectra in (a).
The energy gaps at 10 and 370~K evaluated from Fig.~\ref{fig:oc}(a) are indicated with the solid and open triangles, respectively, and the guide lines for the expected \OC spectra near the energy gaps are shown with dotted lines.
The negative \OC near the energy gap at 10~K is considered an error of the Kramers-Kronig analysis.
(Inset) Wide-range \OC spectrum at $T = 370~{\rm K}$.
}
\label{fig:R}
\end{figure}
Single-crystalline \MnTe samples were synthesized by a chemical vapor transport method, 
and the characterization is described in detail in Sec.~S1 in the supplementary material~\cite{SM}.
The overall character of the grown samples is insulating, as previously reported\cite{Yang2017-aa,Wang2022-aa,Wu2025-aa,Uykur2026-rc}.
The as-grown surface along the $c$-plane was measured for near-normal-incident polarized optical reflectivity [$R(\omega)$] spectra.
The \R spectra were acquired in a wide photon energy \hw range of 8~meV -- 30~eV to ensure accurate Kramers-Kronig analysis (KKA)~\cite{Kimura2013-rg}.
THz-to-visible measurements at \hw = 8~meV -- 2.5~eV have been performed using \R measurement setups at varying temperatures of 10--370~K~\cite{Kimura2008-tg}.
The absolute values of \R spectra were determined with the {\it in-situ} gold-evaporation method.
In the \hw region between 0.1 and 1.4~eV, since the sample is thin (thickness $\sim 0.35$~mm), flat, and transparent, interference of the reflected lights from the front and back surfaces is visible as shown in the inset of Fig.~\ref{fig:R}(a).
To remove interference, we used the multireflection function (Eq.~S1 in the supplementary material) for thin films to derive the essential optical constants, described in Sec.~S2 in the supplementary material~\cite{SM}.
In the \hw range of 1.5--30~eV, the \R spectrum was acquired only at 300~K at the beamline 3B~\cite{Fukui2014-wz} of the UVSOR-III Synchrotron, the Institute for Molecular Science~\cite{Ota2022-ak} for conducting KKA, are shown in the main figure of Fig.~\ref{fig:R}(a).
In order to obtain \OC via KKA of \R, the spectra were extrapolated below 8~meV with a constant reflectivity due to the insulating \R spectra, and above 30~eV with a free-electron approximation ($R \propto \omega^{-4}$)~\cite{Dressel2002-ae}.
It should be noted that magnon absorption appears in the low-energy extrapolation region~\cite{Dzian2025-gv,Uykur2026-rc}, but the contribution to the \R spectra can be neglected due to the low extinction coefficient.
THz reflectivity measurements under negative-strain conditions were performed using the THz micro-spectroscopy beamline (BL6B) at the UVSOR-III Synchrotron~\cite{Kimura2006-mq,Kimura2007-ot}; the experimental setup is described in Sec.~S4 in the supplementary material~\cite{SM}.

Figure~\ref{fig:R}(b) shows the temperature dependence of the \OC spectrum obtained from \R spectra in Fig.~\ref{fig:R}(a).
An optical phonon peak appears at around 17~meV, and an interband transition with an energy gap above 1~eV is observed.
A broad peak, namely an in-gap state, developing at low temperatures was observed in the low-energy region of this energy gap, between 30 and 200~meV.
The following discussion focuses primarily on the temperature dependence of these three structures: the energy gap, the in-gap state, and the optical phonon.
It should be noted that a temperature dependence is observed in the 0.2--1~eV range, with the \OC intensity becoming negative at low temperatures; however, this structure is considered an error of the \R spectrum measurement and of KKA, 
because the intensity of \R spectra derived from expected absorption spectra in the transparent region below the energy gap is less than 0.3-\% difference from that of the experimentally obtained spectra.
Indeed, in the \R spectrum shown in the inset of Fig.~\ref{fig:R}(a), interference between surface and back-surface reflections is observed in this region.
This interference is evidence that the sample is transparent, and the \OC must be nearly zero.
On the other hand, in the \R spectra below 200~meV where the in-gap state is observed at 10~K, the interference disappears as shown in the inset of Fig.~\ref{fig:R}(a), indicating that absorption as well as the \OC intensity appears.
The disappearance of the interference suggests that the in-gap state emerges only at lower temperatures.



\begin{figure}
\includegraphics[width=0.37\textwidth]{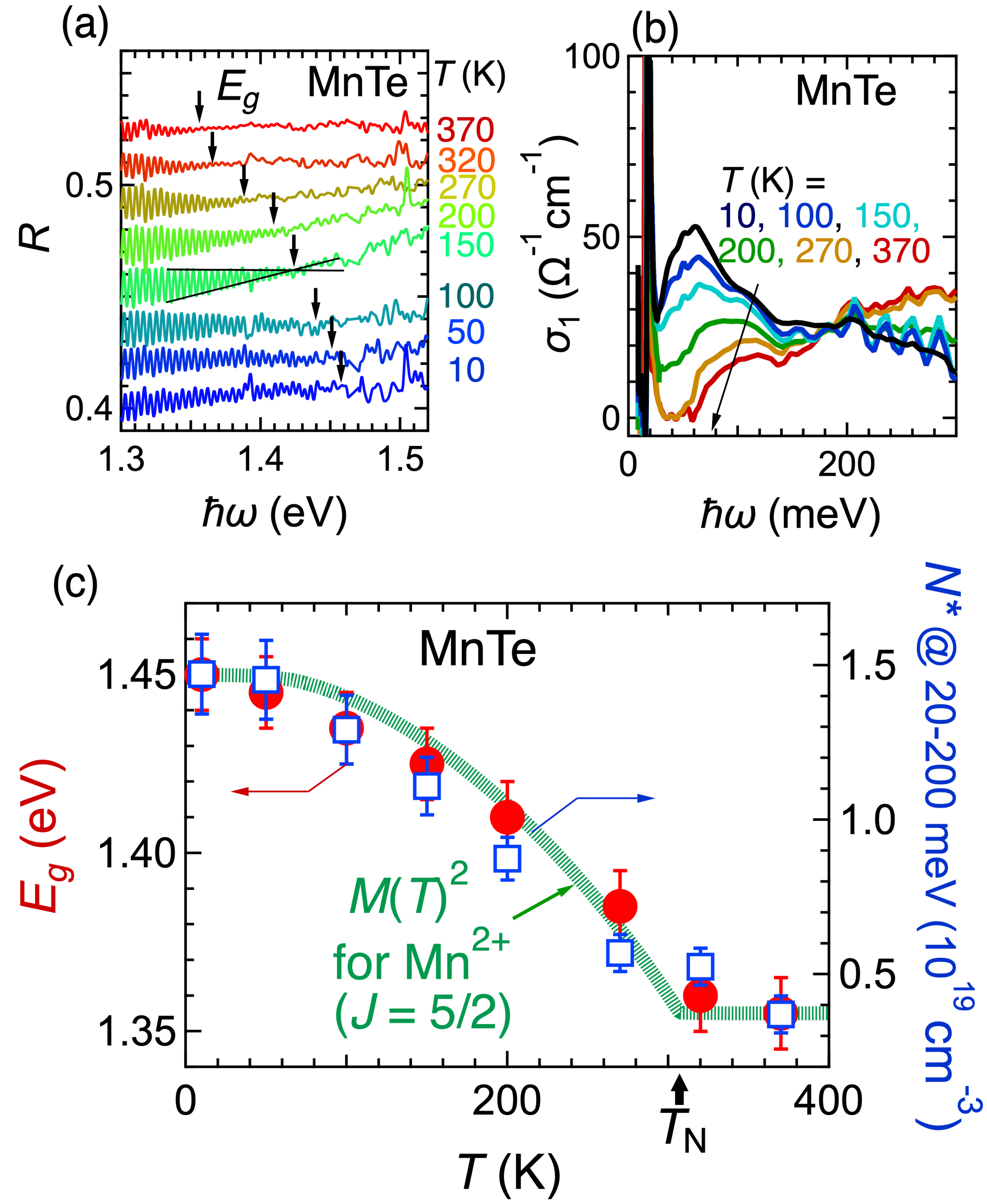}
\caption{
(a) Temperature-dependent \R spectra near the energy gap.
Each spectrum is vertically shifted by 0.02 as the temperature increases.
As shown in Fig.~\ref{fig:R}(a), interferences are visible below the energy gap, and the ends of the interference (marked by down arrows, evaluated as the crossing point of the envelope lines as shown at $T = 150~{\rm K}$) correspond to the onset of the energy gap ($E_g$).
(b) Temperature-dependent \OC spectra in the \hw region below 300~meV.
(c) $E_g$ (red solid circles) and the effective electron number (\Neff, blue open squares) evaluated in the \hw range of 20--200~meV as a function of temperature, accompanied by the square of magnetization [$M(T)^2$, thick dashed line] expected with the Mn$^{2+}~3d^5$ high-spin state ($J = 5/2$).
\TN is shown by a vertical arrow.
}
\label{fig:oc}
\end{figure}
Figure~\ref{fig:oc}(a) shows the temperature dependence of the interference fringes near the energy gap.
These interference fringes arise due to the transparent character of the thin sample, and the point where the fringes disappear (marked by down arrows) can be regarded as the onset of the energy-gap absorption ($E_g$).
The $E_g$ shown by solid circles in Fig.~\ref{fig:oc}(c) increases by about 0.1~eV from about 1.35 to 1.45~eV with decreasing temperature.
This result agrees with previous optical absorption measurements~\cite{Bossini2020-gk}, which claimed proportionality to the square of the magnetization~\cite{Ferrer-Roca2000-fq}.
Indeed, the observed $E_g$ in Fig.~\ref{fig:oc}(c) appears to lie on the square of the magnetization, assuming the high spin of Mn$^{2+}~3d^5$ ($J=5/2$), indicated by the thick dashed line.
$E_g$ following the square of the magnetization has also been observed in the ferromagnetic semiconductor EuO, attributed to band splitting associated with the ferromagnetic transition~\cite{Kimura2008-kh}.
Thus, the temperature dependence of $E_g$ in \MnTe, which possesses an antiferromagnetic spin arrangement with no net magnetism, suggests that this material is altermagnet with band splitting like ferromagnetic materials.

On the other hand, the temperature dependence of the \OC spectrum at around the in-gap state is shown in Fig.~\ref{fig:oc}(b).
Note that a sharp peak at $\sim17$~meV originates from optical phonons.
At $T \geq270~{\rm K}$, a gradual increase toward the high-energy side appears, whereas an in-gap state with a peak at about 60~meV is clearly visible at lower temperatures.
To quantitatively evaluate this spectral change, the effective electron number (\Neff), which corresponds to the integration of the \OC spectra in the \hw range from 20~meV of the phonon peak tail to 200 meV of the point of equal \OC intensity, is shown by the open squares in Fig.~\ref{fig:oc}(c).
The temperature dependence of the open square increases from \TN toward low temperatures, exhibiting a shape nearly identical to that of the solid circles for the energy gap onset, and is also proportional to $M(T)^2$.
This result suggests that the in-gap state also originates from the altermagnetic electronic structure.
Furthermore, the \OC peak intensity above 200~meV at high temperatures shifts to 60~meV at low temperatures, with an estimated energy shift of $\geq140$~meV.
This change is opposite to $E_g$, and the magnitude of the change is not identical.
The expected change in the electronic structure is discussed later.

The observed temperature dependence of the in-gap state is in close agreement with the changes observed in ARPES~\cite{Hajlaoui2024-br}:
The top of the valence band splits into two peaks below \TN.
The shallowest peak shifts from a binding energy of about 200~meV at 300~K to less than 100~meV at 50~K, and its intensity increases with decreasing temperature.
The behavior is similar to that of the in-gap state in the \OC spectrum.
Therefore, the origin of the in-gap state is consistent with the shallowest valence band appearing in ARPES.

\begin{figure}[t]
\includegraphics[width=0.4\textwidth]{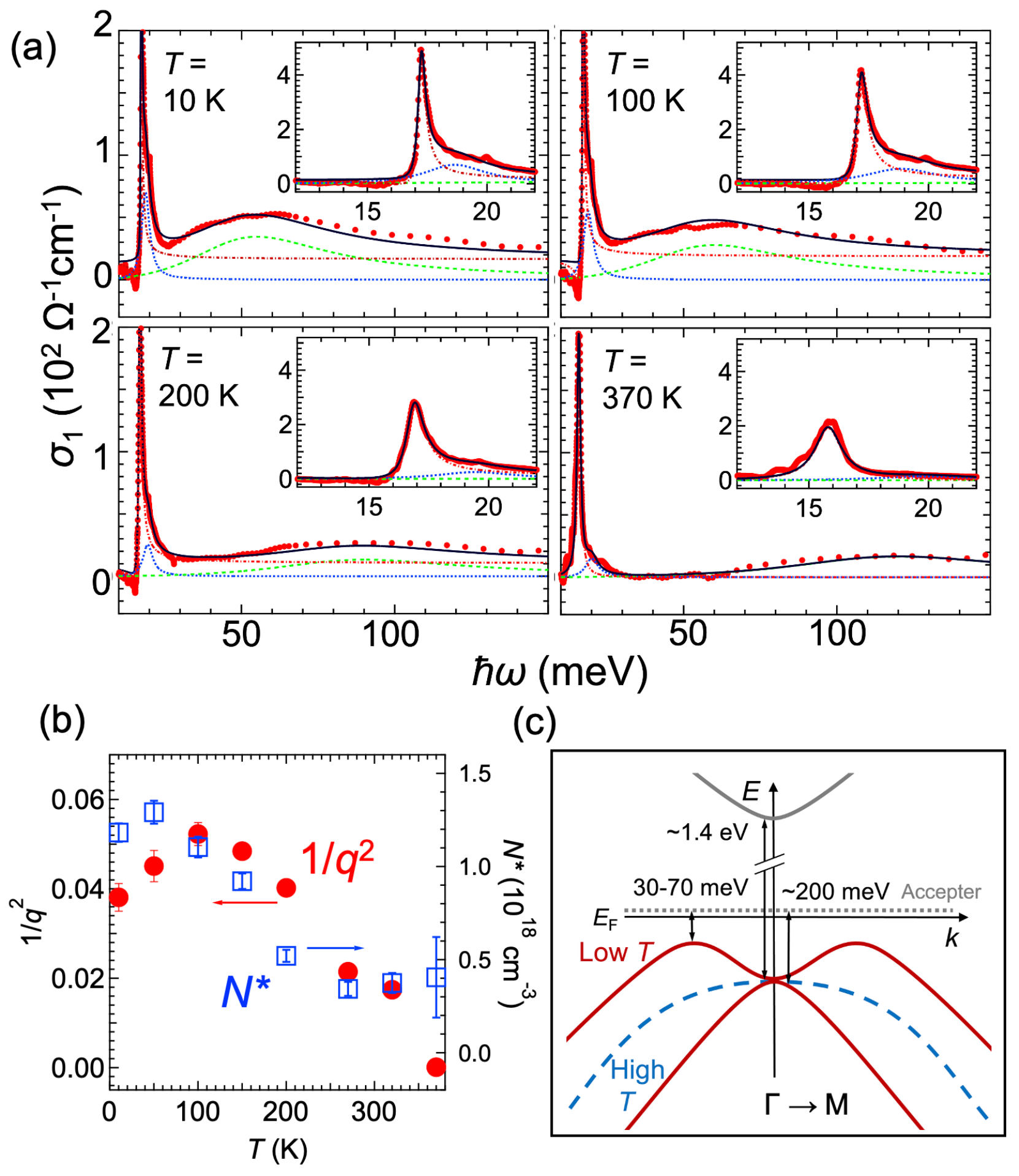}
\caption{
(a) \OC spectra (circles) at temperatures of 10, 100, 200, and 370~K in the \hw region below 150~meV and the fitting curves with three peaks at about 17, 20, and 60~meV at 10~K.
(b) Fano asymmetric parameter ($1/q^2$, red solid circles) of the phonon peak and the peak intensity (\Neff, blue open squares) of the 20-meV peak as a function of temperature.
(c) Expected temperature-dependent electronic-structure change near \EF and the origins of the peaks in \OC spectra.
}
\label{fig:phonon}
\end{figure}
Next, we discuss the relationship between the in-gap state and the TO phonon (Mn-Te stretching mode).
Figure~\ref{fig:phonon}(a) shows the \OC spectrum in the low-energy region at representative temperatures (Other temperatures' \OC spectra and fitting results are shown in Fig.~S2(a) in the supplementary material~\cite{SM}).
The structure consists of three peaks: 
A sharp peak at about 17~meV originating from optical phonons, a slightly broad peak centered at around 20~meV, and a broad peak above 30~meV, as previously described as an in-gap state, as shown in Fig.~\ref{fig:oc}(b).
The phonon peak has been observed in previous papers~\cite{Allen1977-ae} and explained with phonon calculations~\cite{Mu2019-be}.
The 20-meV peak is not very strong, but it shows a clear temperature dependence in both intensity and peak energy.
Around this energy, there is a large phonon density of states in calculations~\cite{Mu2019-be}; therefore, a phonon absorption is plausible for the origin.

The 17-meV phonon peak exhibits a Fano-like asymmetric line shape~\cite{Fano1961-kl,Miroshnichenko2010-fx}, suggesting a strong coupling between the Mn-Te stretching and the electronic background of the in-gap state~\cite{Lupi1998-ii,Kimura2007-wz}.
The other two peaks are broad and therefore symmetric, so the spectra were fitted with one Fano and two Lorentzian functions as shown in Eq.~S2 in the supplementary material~\cite{SM}.
The Fano asymmetric parameter ($1/q^2$) for the phonon peak and the intensity of the 20-meV peak are shown in Fig.~\ref{fig:phonon}(b), and other obtained fitting parameters are shown in Fig.~S2(b-d) in the supplementary material~\cite{SM}.
Note that another optical Lorentz model with the Fano effect was described in previous papers~\cite{Joe2006-af,Gallinet2011-ie,Wenzel2022-rb}, but the temperature dependence of the fitting parameters was consistent with those obtained from the original Fano line shape.

The results revealed that the $1/q^2$ Fano term increased markedly from 0 to 0.05 as the temperature decreased from 370 to 100~K.
The temperature dependence of $1/q^2$ is similar to the intensity of the in-gap state shown in Fig.~\ref{fig:oc}(c), even though $1/q^2$ decreases from 100~K to 10~K.
Therefore, the in-gap state might affect the Fano asymmetric parameter $1/q^2$.
The Fano effect originates from the interaction between a localized peak and a continuous background~\cite{Fano1961-kl}.
In the present case, a possible continuous background is absorption in the in-gap state, as there is no Drude background due to the absence of carriers.
Since the in-gap state is thought to originate from the altermagnetic spin-split bands, as discussed above, the appearance of the altermagnetic electronic structure might influence the Mn-Te stretching mode.

Based on the results and ARPES data, the observed \OC spectra reflect the electronic structure as shown in Fig.~\ref{fig:phonon}(c).
The main energy gap is $\sim1.4$-eV wide as shown in Fig.~\ref{fig:oc}(c), and the gap size increases about 0.1~eV with decreasing temperature, suggesting the bottom of the conduction band shifts to the high energy side by 0.1~eV.
On the other hand, the in-gap state is related to the spin-split band, which has a similar temperature dependence observed in ARPES~\cite{Hajlaoui2024-br}, {\it i.e.}, the top of the valence band has no spin splitting above \TN and lies approximately 0.2 eV below \EF, but below \TN, one of the spin-split bands approaches \EF.
Its energy shift is about 0.15~eV, which closely matches the energy shift observed for the in-gap state in this study.
ARPES data also suggest that \MnTe is a $p$-type semiconductor~\cite{Krempasky2024-ew}, so an acceptor level is located just above \EF.
The in-gap state is considered to originate from the transition from the spin-split band to the acceptor level.

\begin{figure}[t]
\includegraphics[width=0.4\textwidth]{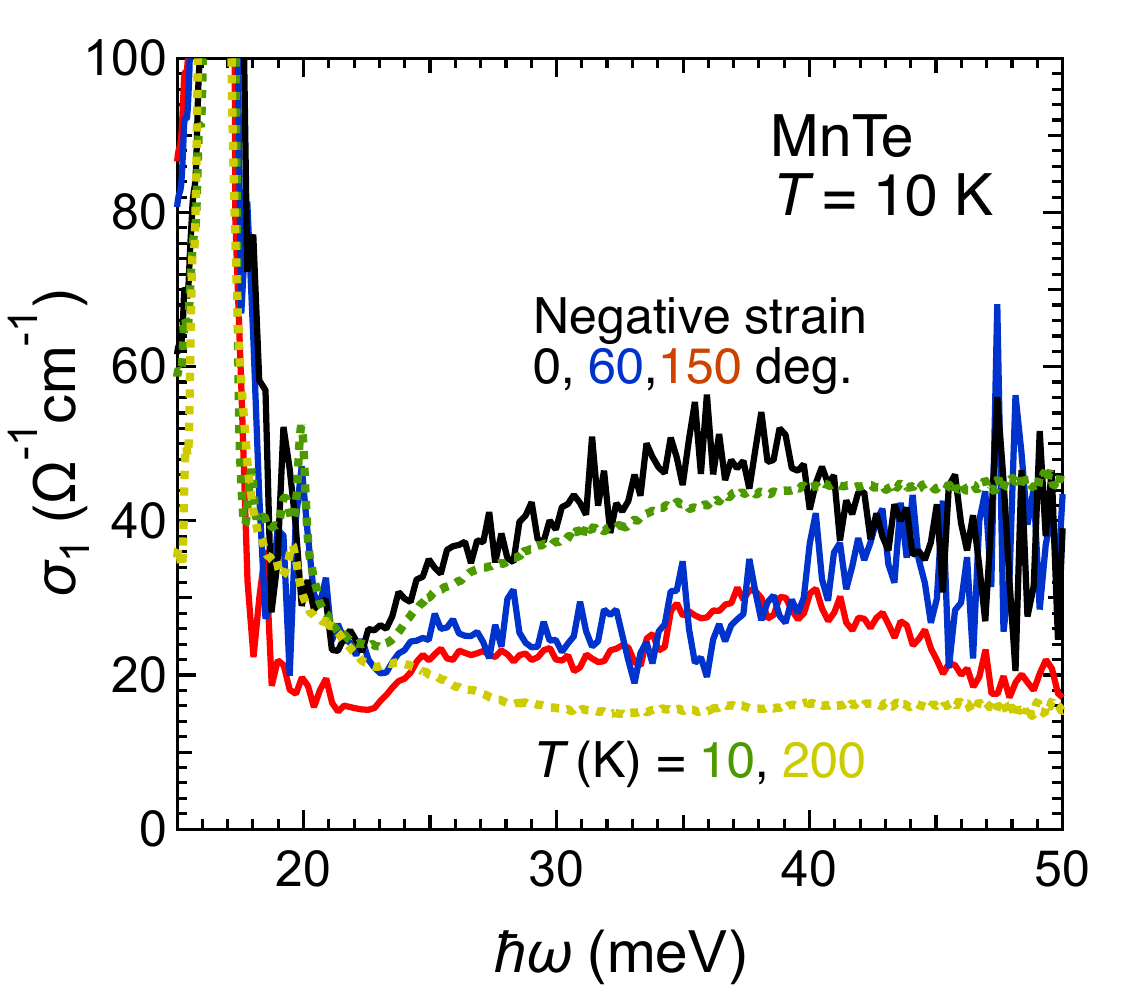}
\caption{
Negative-strain dependence of the \OC spectrum of \MnTe at 10~K (solid lines) compared with the spectra at 10 and 200~K without applying strain (dashed lines).
The value of the negative strain is described by the applied angle of the screw rotation, which is proportional to the negative strain strength with $\sim 0.8~\%$ at 360~degrees without samples, as shown in Fig.~S3(c) in the supplementary material~\cite{SM}.
}
\label{fig:pdep}
\end{figure}
Finally, we investigated the theoretically predicted strain effect on the spin-split band~\cite {Belashchenko2025-mm}.
To apply a strain less than 1~\% and to put it at the cold finger of a conventional cryostat, we introduced a mechanical uniaxial cell designed with reference to Ref.~\cite{Ricco2018-co} as shown in Fig.~S3 in the supplementary material~\cite{SM}.
Figure~\ref{fig:pdep} shows the negative strain dependence of the \OC spectrum at \hw $\leq 50~{\rm meV}$ and a temperature of 10~K (Original \R spectra are shown in Fig.~S4 in the supplementary material~\cite{SM}).
The figure shows that the in-gap peak at around 35~meV is strongly suppressed on the lower-energy side from 0 to 60~degrees of push screw rotation for applying negative strain, and is smeared out at 150 degrees.
The spectral change is similar to the effect of heating from 10 to 200~K, in which the top of the spin-split band moves away from \EF, meaning the spin-splitting width decreases.
This suggests that the splitting width of the spin-split bands appearing near \EF becomes smaller under negative strain.
This result corresponds to the theoretically predicted decrease in the spin-splitting angle $\theta_{SS}$ under negative strain of less than 1~\%~\cite{Belashchenko2025-mm}.


To summarize, we have investigated temperature- and strain-dependent modifications in the electronic structure of an altermagnet candidate, \MnTe.
In addition to the energy gap structure at \hw of $\sim1.4$~eV, we found an in-gap state in the THz region.
The in-gap state has a strong temperature dependence, in which the spin-split band appears and approaches \EF below \TN, which is consistent with the change in the electronic structure of an altermagnet.
The in-gap state also exhibits a strong negative strain dependence, as theoretically predicted.
The temperature and negative-strain dependence suggest that the spin-split band can be controlled with such perturbations.
It was also observed that the optical phonon of the Mn-Te stretching mode exhibits a strong Fano effect, originating from a continuous background, likely from the in-gap state.


\section*{Acknowledgments}
We would like to thank the staff members of the UVSOR Synchrotron Facility for their support during synchrotron radiation experiments and the staff members of the Equipment Development Center of the Institute for Molecular Science for prototyping the strain cell.
Part of this work was performed under the Use-of-UVSOR Synchrotron Facility Program (Proposals Nos.~24IMS6018, 24IMS6024, 25IMS6014, 25IMS6022) of the Institute for Molecular Science, National Institutes of Natural Sciences.
This work was partly supported by JSPS KAKENHI (Grant Nos.~24K21197, 23H00090, 23H00184), the National Research Foundation of Korea (NRF) (Grant Nos.~RS202500516322, RS202500512822, and RS2022NR068225), and the National Research Council of Science \& Technology (NST) grant by the Korea government (MIST) (Grant No.~GTL 25091-102).


\vspace{5mm}

{\it Note added}: During completing this manuscript, we became aware of a related work~\cite{Gao-Gao-2026-yo}. 
We found that their energy gap near the energy gap is inconsistent with the previous works~\cite{Zanmarchi1967-nm, Allen1977-ae, Angadi1994-nl, Janik1996-dl, Ferrer-Roca2000-fq, Mori2018-lo, Bossini2020-gk}.
This might be due to errors arising from the absence of high-energy spectral data, even though the KKA method requires spectra spanning a very wide energy range.
In addition, their \OC data are inconsistent with ours, including the absence of an in-gap state and the emergence of a Drude component.

%
%

\bibliography{MnTeOptics.bib}

\end{document}



\title{
Supplementary material for ``Control of band structure in the altermagnetic candidate MnTe by temperature and strain''
}
    
\author{Shin-ichi~Kimura}
\email{sk@kimura-lab.com}
\affiliation{Graduate School of Frontier Biosciences, The University of Osaka, Suita, Osaka 565-0871, Japan}
\affiliation{Department of Physics, Graduate School of Science, The University of Osaka, Toyonaka, Osaka 560-0043, Japan}
\affiliation{Institute for Molecular Science, Okazaki, Aichi 444-8585, Japan}
\author{Hironao~Suwa}
\affiliation{Department of Physics, Graduate School of Science, The University of Osaka, Toyonaka, Osaka 560-0043, Japan}
\author{Kangle~Yuan}
 \altaffiliation[Present affiliation: ]{The Institute for Solid State Physics, The University of Tokyo, Kashiwa, Chiba 277-8581, Japan}
\affiliation{Department of Physics, Graduate School of Science, The University of Osaka, Toyonaka, Osaka 560-0043, Japan}
\author{Hiroshi~Watanabe}
 \altaffiliation[Present affiliation: ]{Institute for Chemical Research, Kyoto University, Uji, Kyoto 611-0011, Japan}
\affiliation{Graduate School of Frontier Biosciences, The University of Osaka, Suita, Osaka 565-0871, Japan}
\affiliation{Department of Physics, Graduate School of Science, The University of Osaka, Toyonaka, Osaka 560-0043, Japan}
\author{Takuto~Nakamura}
\affiliation{Graduate School of Frontier Biosciences, The University of Osaka, Suita, Osaka 565-0871, Japan}
\affiliation{Department of Physics, Graduate School of Science, The University of Osaka, Toyonaka, Osaka 560-0043, Japan}
\author{Haan Kyul Yun}
\affiliation{Department of Physics, Sogang University, Seoul 04107, Republic of Korea}
\author{Myung-Hwa~Jung}
\affiliation{Department of Physics, Sogang University, Seoul 04107, Republic of Korea}

%

%
\maketitle
\setcounter{equation}{0}
\setcounter{figure}{0}
\setcounter{table}{0}
\setcounter{section}{0}
\setcounter{page}{1}
\makeatletter
\renewcommand{\thesection}{S\arabic{section}}
\renewcommand{\theequation}{S\arabic{equation}}
\renewcommand{\thefigure}{S\arabic{figure}}
\renewcommand{\bibnumfmt}[1]{[S#1]}
\renewcommand{\citenumfont}[1]{S#1}

\section{Single crystal growth and characterization}

Single-crystalline \MnTe samples were synthesized by a chemical vapor transport method.
First, a stoichiometric mixture of high-purity Mn powder and Te shot was evacuated and sealed in a quartz ampoule.
The mixture was placed at $500~^\circ{\rm C}$ for a solid-state reaction to synthesize homogeneous polycrystalline \MnTe.
The reacted polycrystal was then placed in a two-zone furnace with iodine (I$_2$) as the transport agent.
The temperatures were $700~^\circ{\rm C}$ for the source zone and $685~^\circ{\rm C}$ for the sink zone.
After 14~days, $p$-type single-crystalline samples with a thickness of about $350~\mu{\rm m}$ were obtained.

\begin{figure*}[t]
\includegraphics[width=0.95\textwidth]{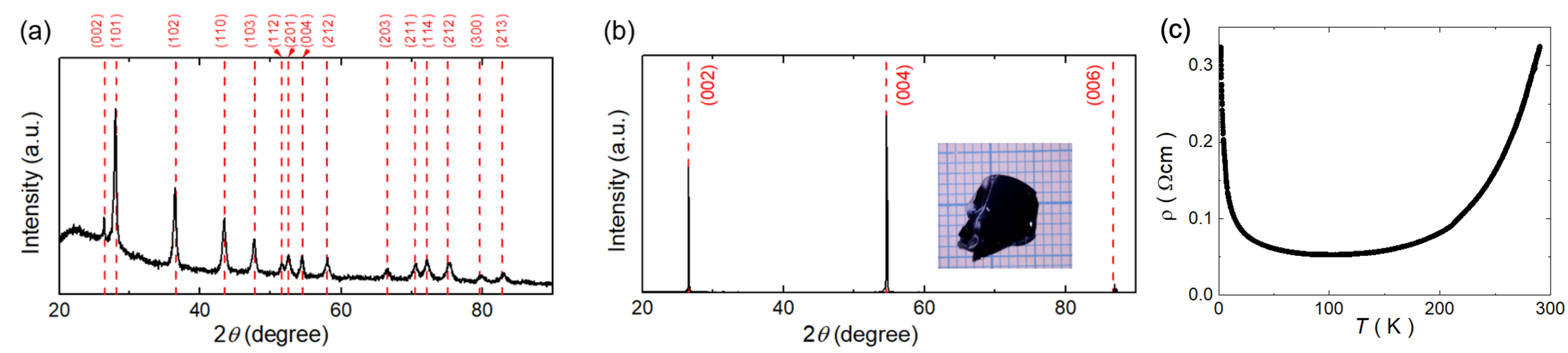}
\caption{
(a) Powder XRD pattern indexed by hexagonal $\alpha$-MnTe.
(b) Single-crystal XRD pattern showing the expected (001) reflections. The inset shows an optical image of representative plate-like crystals.
(c) Temperature dependence of electrical resistivity, $\rho(T)$, of our MnTe sample, showing nonmetallic transport behavior.
}
\label{fig:supfig0}
\end{figure*}
The powder X-ray diffraction (XRD) pattern shown in Fig.~\ref{fig:supfig0}(a) is well indexed to the hexagonal $\alpha$-MnTe phase, without detectable impurity peaks from MnTe$_2$ or other Mn-Te secondary phases.
The single-crystal XRD pattern in Fig.~\ref{fig:supfig0}(b) shows only the expected (001) reflections, confirming the well $c$-axis-oriented single-crystalline character of the plate-like crystal.
Furthermore, the temperature-dependent resistivity in Fig. ~\ref{fig:supfig0}(c) shows nonmetallic transport behavior, with a broad minimum and a low-temperature upturn.
This behavior is consistent with previously reported MnTe transport data~\cite{Yang2017-aa,Wang2022-aa,Wu2025-aa,Uykur2026-rc} and is distinct from a strongly doped case~\cite{Gao-Gao-2026-yo}.
These structural and transport results indicate that the measured optical response is not dominated by MnTe$_2$ secondary phases or heavily doped impurity contributions.

\section{Removing interferences from reflectivity spectra}

To remove interference from $R_{obs}(\omega)$ spectra, the spectra were fitted with the function 
\begin{eqnarray}
R_{obs}(\omega) &=& \left|\dfrac{R(\omega)^{1/2}(1+\exp \{i \phi(\omega) - \beta(\omega)\})}{1+R(\omega) \exp \{i \phi(\omega) - \beta(\omega)\}}\right|^2, \nonumber \\
\phi(\omega) &=& \dfrac{n(\omega) \omega \delta}{c},  \beta(\omega) = \dfrac{2 \kappa(\omega) \omega \delta}{c}, \nonumber \\
R(\omega) &=& \dfrac{\{1-n(\omega)\}^2+\kappa(\omega)^2}{\{1+n(\omega)\}^2+\kappa(\omega)^2}
\end{eqnarray}
Here, $R_{obs}(\omega)$ and $R(\omega)$ are the observed and actual reflectivity spectra, $n(\omega)$ and $\kappa(\omega)$ are the refractive index and extinction coefficient, $\delta$ and $c$ are the sample thickness and light velocity, respectively, and $\omega=2\pi\nu$, where $\nu$ is frequency (wavenumber).
The fitting cycle contains the Kramers-Kronig relation of $n(\omega)$ and $\kappa(\omega)$. 

\section{Fitting THz spectra with Fano and Lorentz functions}

\begin{figure*}[t]
\includegraphics[width=0.8\textwidth]{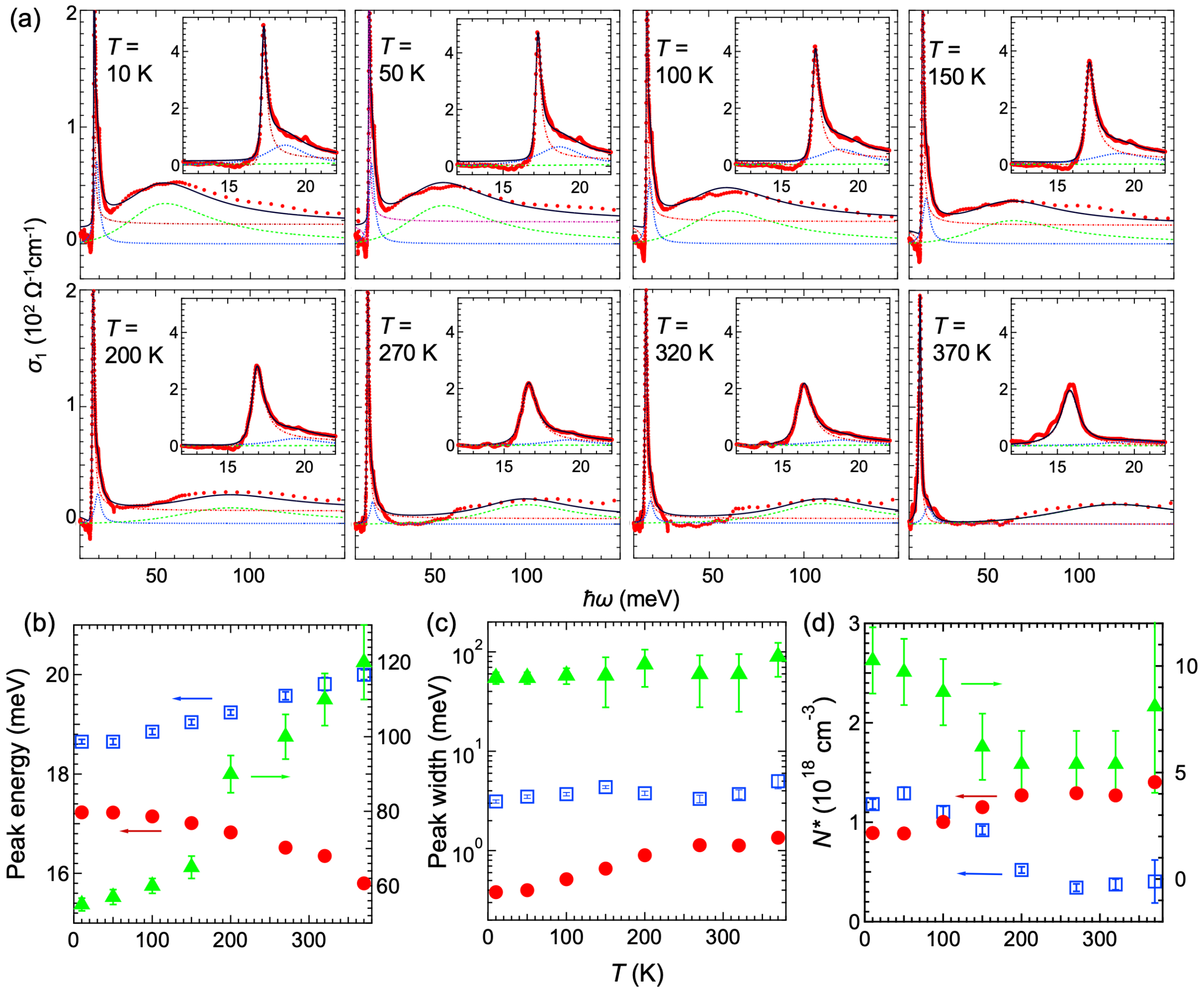}
\caption{
(a) \OC spectra (red circles) of all measured temperatures in the \hw range below 150~meV and fitted by eq.~\ref{eq:fitting}.
The Fano function fitted to the phonon line, the Lorentz function fitted to the 20-meV peak, and the Lorentz function fitted to the in-gap peak are plotted as magenta dot-dashed, blue dotted, and green dashed lines, respectively, and their sum is shown as a black solid line.
(Inset) Enlarged figure in the \hw range of 12--22~meV.
Peak energies (b), peak width (c), and effective electron number (\Neff, d) of three peaks as a function of temperature.
The phonon peak, the 20-meV peak, and the in-gap peak are plotted as red solid circles, blue open squares, and green solid triangles, respectively.
}
\label{fig:supfig1}
\end{figure*}
To obtain temperature-dependent parameters in the THz region shown in Fig.~\ref{fig:supfig1}, we used the following function:
\begin{equation}
\displaystyle \sigma_1(\omega) = \dfrac{N_{ph}^* e^2}{2 m_0 q^2 \Gamma_{ph}} \dfrac{(\varepsilon+q)^2}{\varepsilon^2+1}
 + \sum_{i=1}^2 \dfrac{N_{i}^* e^2}{m_0} \dfrac{\Gamma_i \omega^2}{(\omega_i^2-\omega^2)^2 + \Gamma_i^2 \omega^2}
\label{eq:fitting}
\end{equation}

The first term describes the asymmetric Fano line shape of the very sharp phonon peak appearing at $\sim17$~meV, where $\varepsilon=\dfrac{\omega-\omega_{ph}}{\Gamma_{ph}/2}$ is a reduced energy, $\omega_{ph}$, $\Gamma_{ph}$, $N_{ph}^*$, and $q$ are the peak energy, the width, the effective electron number, and the asymmetric paramter of the phonon, respectively~\cite{Fano1961-kl,Miroshnichenko2010-fx}.
$e$ and $m_0$ are the charge and the rest mass of the electron, respectively.
The latter term corresponds to the broad peaks at $\sim20$~meV for $n=1$ and $\sim50$~meV for $n=2$, where $\omega_n$, $\Gamma_n$, and $N_n^*$ are the peak energy, the width, and the effective electron number, respectively.
The latter shape arises from the optical Lorentz model~\cite{Dressel2002-ae}, which is inconsistent with the ``general'' Lorentz function described as the symmetry limit ($|q| \to \infty$) of the Fano line shape.
In the case of a narrow peak, such as the 17-meV phonon peak with a large $q$ value, the peak shape is very similar to the optical Lorentz function; but when the width $\Gamma$ is close to the peak energy in a broad peak case, the shape is markedly different, especially the intensity at $\omega = 0$ appears even though the insulating character.
Then, we adopted the mixed function~\ref{eq:fitting} for the fitting.
Fitting results for all temperatures are shown in Fig.~\ref{fig:supfig1}.
Note that an optical Lorentz model with the Fano effect was described in previous papers~\cite{Joe2006-af,Gallinet2011-ie}, but the temperature dependence of the relation of asymmetry of the phonon shape to the altermagnetic spin-split band can be derived from the simple function~\ref{eq:fitting}.
Additionally, we also fitted using another Fano-like Lorentz function:
\begin{equation}
\displaystyle \sigma_1(\omega) = \sum_{i=0}^2 \dfrac{\sigma_i \omega \Gamma_i}{4 \pi} \dfrac{\Gamma_i \omega (q_i^2-1)+2 q_i (\omega^2-\omega_i^2)}{(\omega^2-\omega_i^2)^2 + \Gamma_i^2 \omega^2}
\label{eq:winzel_fitting}
\end{equation}
described by Winzel {\it et al.}\cite{Wenzel2022-rb}.
The fitting results are almost consistent with Fig.~\ref{fig:supfig1} because of the very narrow phonon peak and the symmetric Lorentz shapes of peaks $i=1$ and $2$.

\section{Uniaxial negative strain cell}

\begin{figure}
\includegraphics[width=0.45\textwidth]{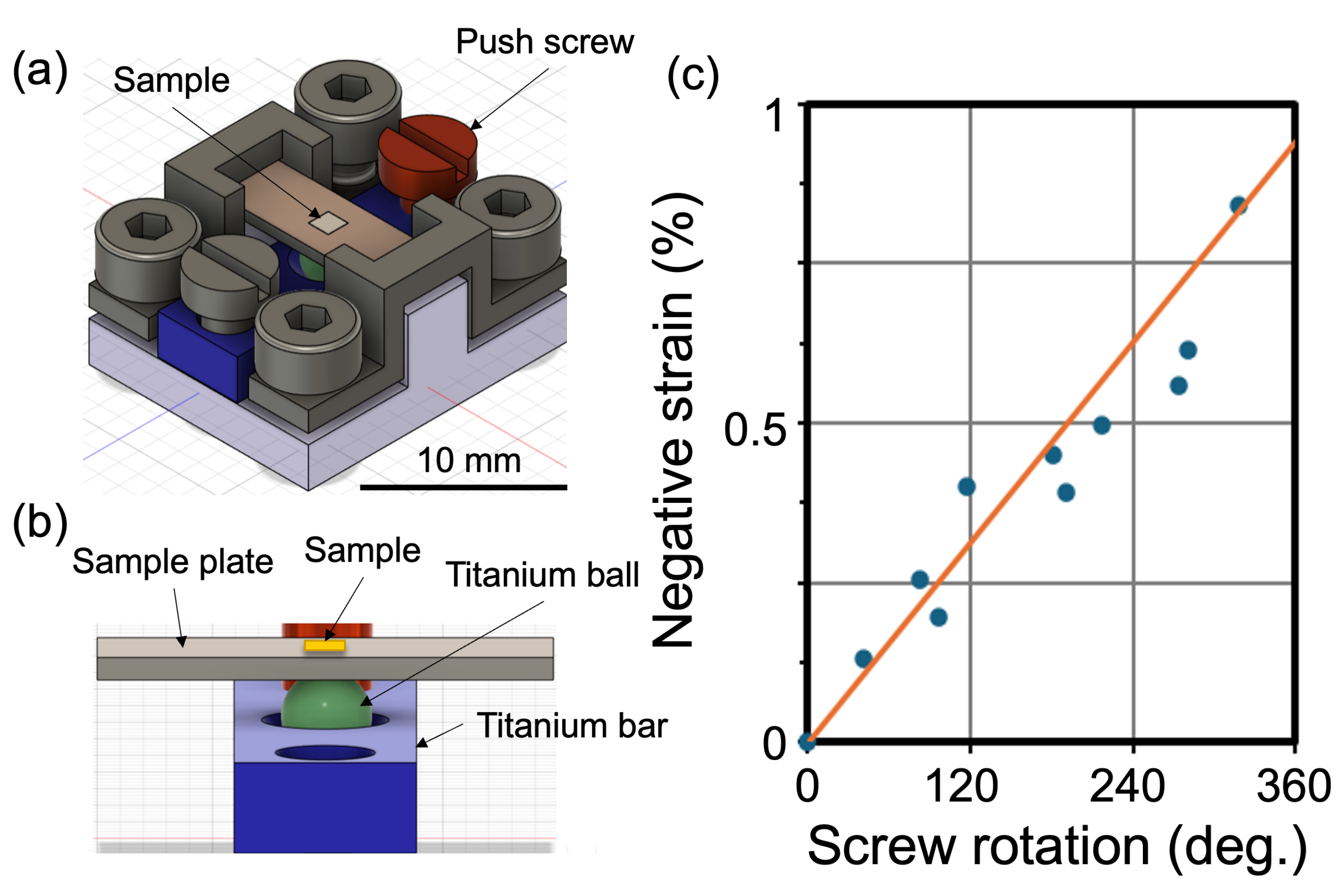}
\caption{
(a) The developed negative strain cell.
Rotating the flat-head screw (red) pushes the titanium rod (blue) upward, which in turn pushes the titanium ball (green) upward, applying strain to the copper substrate with the sample attached.
(b) Enlarged view of the strain-applying titanium rod (blue), the strain-applying titanium ball (green), and the copper substrate for attaching the sample.
(c) Experimental results (circles) and a simulated result (solid line) of negative strain versus screw rotation angle using strain gauges. 
Experimentally obtained negative strain values varied linearly with screw rotation angle, yielding approximately 0.8~\% strain over a 360-degree rotation.
}
\label{fig:supfig2}
\end{figure}
Here, we prototyped a uniaxial negative-stress cell for infrared and THz measurements, as shown in Fig.~\ref{fig:supfig2}(a), which was designed with reference to \cite{Ricco2018-co}.
The cell size is approximately $10({\rm W})\times12({\rm L})\times5({\rm H})$~mm$^{3}$ for installation in a cryostat (MicrostatHe, Oxford Instruments).
The stress is controlled by a screw that raises a titanium bar to press a titanium ball located beneath a sample bar made from a 0.5-mm-thick copper plate (Fig.~\ref{fig:supfig2}(b)).
The magnitude of the strain was evaluated with a strain gauge (C5K-06-S5145-350-33F, Micro-Measurements), which is consistent with the simulation result using CAD software (Fusion 360, Autodesk Inc.), as shown in Fig.~\ref{fig:supfig2}(c).
In Fig.~4, the magnitude of the strain is expressed by the screw's rotation angle because it is changed by the attached samples.

Since the size of the almost flat stress area on the sample bar is smaller than $0.5\times0.5$~mm$^2$, \MnTe samples were cut to a size less than the flat area.
To measure reflectivity spectra of such small samples, we used the synchrotron THz micro-spectroscopy beamline~\cite{Kimura2006-mq,Kimura2007-ot} at the UVSOR Synchrotron~\cite{Ota2022-ak}.
The THz light is horizontally polarized and nearly aligned with the [1000] direction and the negative strain direction of the uniaxial cell.
\\
\section{Reflectivity spectra at negative strain}

\begin{figure}
\includegraphics[width=0.45\textwidth]{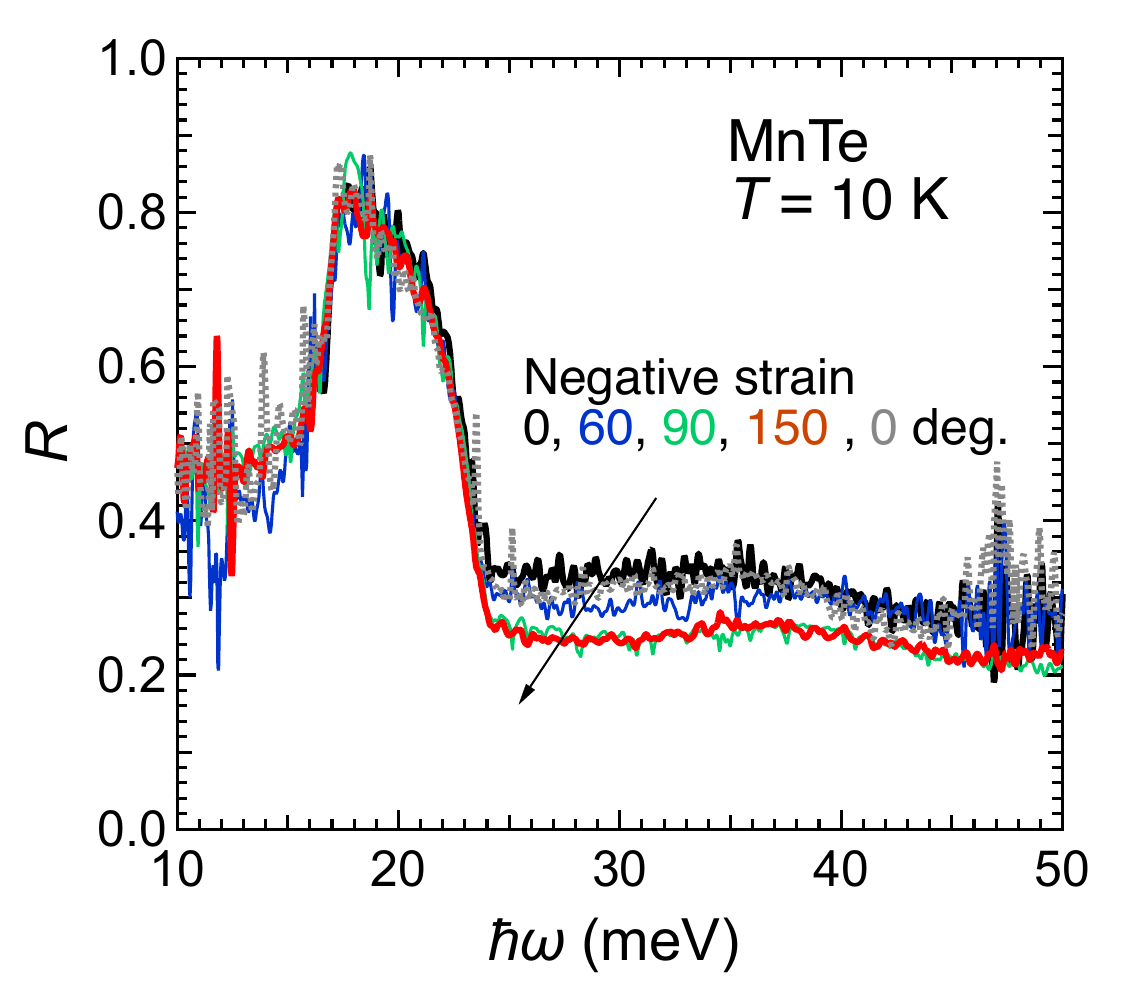}
\caption{
Negative-strain-dependent \R spectra of \MnTe at $T = 10~{\rm K}$.
The magnitude of the negative stress is described by the rotation angle of the screw.
The gray dotted line represents the spectrum after the stress is released.
}
\label{fig:supfig3}
\end{figure}

Figure~\ref{fig:supfig3} is the negative-strain-dependent \R spectra of MnTe at $T = 10~{\rm K}$.
The intensity in the \hw energy region above 25~meV decreases as the strain becomes more negative.
After connecting the higher energy \R spectra without strain, the \OC spectra shown in Fig.~4 were derived from the Kramers-Kronig analysis.

%
%

\bibliography{MnTeOptics.bib}